\documentstyle[12pt,epsf]{article}
\font\tenbf=cmbx10
\font\tenrm=cmr10
\font\tenit=cmti10

\input{psfig.sty}
\begin{document}

\font\fortssbx=cmssbx10 scaled \magstep2
\hbox to \hsize{
\hfill$\vcenter{\hbox{ . }
 \hbox{ . }}$ }

\vspace{.1in}

\begin{center}
{\tenbf Evidence for Compact Dark Matter in Galactic Halos }
\\
\vskip 0.7cm
{\tenrm S.~Abbas }
\\[.1cm]
{ \tenit Department of Physics, Utkal University, Bhubaneswar-751004,
Orissa, India, \\
email: abbas@iopb.res.in }
\\[.5cm]
{\tenrm A.~Abbas}\\
{\tenit Institute of Physics, Bhubaneswar-751005, India, \\
email: afsar@iopb.res.in }
\\[.5cm]
{\tenrm S.~Mohanty}\\
{\tenit Department of Physics, Utkal University, Bhubaneswar-751004,
Orissa, India }
\end{center}

\smallskip

{\footnotesize

\begin{center}ABSTRACT \end{center}
{\narrower

Clumped dark matter arises naturally within the framwork of
generic cosmological
dark matter models. Invoking the existence of dark matter clumps can also
solve may unexplained mysteries in astrophysics and geology or geophysics 
, eg. the
galactic gamma-ray halo and the periodic
terrestrial flood basalt volcanic episodes. 
Clumped dark matter is dynamically stable to
friction and will not heat the disk. Such clumps may have already been
discovered in the form of dwarf
spheroidals, and further searches are encouraged by the results of this
paper.

}}

\section{Clumping as a Prediction of Cosmological Models}

As the bulk of the matter in the Universe is dark, a better understanding
through theoretical models is important, and much current activity is
taking place in that direction. 
Undoubtedly more significant than this would
be the detection of this dark matter. Much progress is taking place, and
it is possible that dark
matter may have been already detected. A Gran Sasso group claims
to have found anomalous signs of a 56 GeV dark matter candidate
\cite{60}. 
For recent discussions of the DAMA results, consult \cite{belli}, 
\cite{bdfs}.
However, a definite confirmation of this is awaited.

 As the visible matter clumps together to form stars, planets, etc. an
interesting question is whether the dark matter also displays this
tendency
of clumping. Interestingly several dark matter models do
suggest that clumps of dark matter arise naturally during the course of
evolution of the universe.  Silk and Stebbins \cite{silk} considered cold
dark matter models with cosmic strings and textures appropriate for galaxy
formation.  They found that a fraction $ 10^{-3} $ of the galactic halo
dark matter may exist in the form of dense cores. These may survive up to
mass scales of $ 10^8 M_{\odot} $ in galaxy halos and globular clusters
\cite{silk}.  Analysing the stability of these clumps of dark matter, they
found that the cores of these clumps will not be affected, although the
outer layers may be stripped off by tidal forces. In the cosmic string
model, the clumpiness C, defined as the ratio of clumped matter
concentration to normal concentration, of dark matter at the present epoch
would be \cite{silk}

\begin{equation}
C \sim 10^{12} f_{cl} h^6 \Omega_{0}^3
\end{equation}

where $ f_{cl} $ is the fraction of dark matter in clumps, 
$ H $ is the Hubble parameter parametrised as 
$ 100 ~h ~km/s ~Mpc^{-1} $, and 
$ \Omega_{0} $ is the closure energy density of the Universe.  

Subsequently
Kolb and Tkachev \cite{kolb} studied isothermal fluctuations in the dark
matter density during the early universe. If the density of the isothermal
dark matter fluctuation or clumps, 
$ \Phi = \delta \rho_{DM} / \rho_{DM} $,
exceeds unity, a fluctuation collapses in the radiation-dominated epoch
and produces a dense dark matter object. They found the final density
of the virialized object $ \rho_{F} $ to be 

\begin{equation}
\rho_{F} \sim 140 \Phi^{3} ( \Phi + 1 ) \rho_{eq}
\end{equation}

where $ \rho_{eq} $ is equilibrium density.

For axions, a putative dark matter particle,
density fluctuations can be very high, possibly spanning the
range $ 1 < \Phi < 10^4 $. The resultant density in miniclusters can be as
much as $ 10^{10} $ times larger than the local galactic halo density. The
probablility at present of an encounter of the earth with such an axion
minicluster is 1 per $ 10^7 $ years with $ \Phi = 1 $. Kolb and Tkachev
found two types of axion clumps arising from two kinds of initial
perturbations:

\begin{itemize}
 \item{ Fluctuations with $ 10^{-3} < \Phi < 1 $ collapse in the
matter-dominated epoch. }
 \item{ Fluctuations with $ \Phi > 1 $ collapse in the radiation-dominated
epoch. }
\end{itemize}

  If the dark halo is mostly made of neutralinos, then the clumping factor
in the MSSM could be less than $ 10^9 $ for all neutralino masses
\cite{bergstrom}. 


Structure models involving the hierarchical accretion and mergeer of dark
matter halos are well-motivated and attractive. Numerical simulations of
the substructure with galactic and cluster halos which form within
hierarchical universes are easily reproduced, yielding thousands of
substructure clumps. The models analsed by B.Moore, S.Ghighna et al.
\cite{ben} 
predict that the Milky Way should contain at least 500 satellites of bound
masses $ \geq 10^8 M_{\odot} $ and tidally limited sizes $ \geq kpc $.  
Remarkably, dark matter substructure survives on galactic scales, so that
galaxy halos appear as scaled versions of galaxy clusters. One model
considered by S.Ghigna, Ben Moore et al. \cite{ghigna}
resolved 150 halos within a
model galaxy cluster. Sazhin et al. \cite{sazhin} have suggested that 
microlensing events observed towards the Large Magellanic Cloud may
consist of dark matter clumps.

\section{ Astrophysical Constraints
and Stability of Clumps}

Astrophysical constraints can be used to place bounds on the mass of the
clumps. The clumps are also stable with regard to main astrophysical
disruption processes. 

\subsection{\it Disk `Puffing' by Dark Clumps }

As the dark matter clumps traverse the Galactic halo, they will impart
energy to the stars residing there. This leads to a gradual puffing up of
the disk, and the stars present are heated. In this process, the older
stars get heated more than the younger ones. Lacy and Ostriker considered
the case of black holes of mass $ 10^6 M_{\odot} $ generating the
observed
puffing of the galaxy, and concluded that this was the best mechanism to
explain the observed amount of `puffing' of the galactic disk
\cite{carr}. However,
subsequent data revealed that the velocity dispersion of disk stars, $
\sigma $, may no longer rise as quickly as $ \sqrt{t} $. which was one of 
Lacy and Ostricker's assumptions. In addition,
other sources of heat like spiral density waves and giant molecular clouds
may give a better fit to the data. Hence the Lacy-Ostriker model of black
holes generating the observed puffing of the disk is not as convincing now
as it was when first suggested. However, they obtained an upper limit on
the density of halo objects of mass M, which still hold :

\begin{eqnarray}
\Omega_{B} \leq \Omega_{h} ~min \left[ 1, (M/M_{heat})^{-1} \right] \\
{\sf where } ,\,\
M_{heat} = 3 \times 10^6 \left( t_g / 10^{10} \right)^{-1} 
\end{eqnarray}

\noindent where $ t_g $ is the age of the galaxy. This condition must
hold, 
otherwise the disk would be more puffed up than observed. 
Using this formula, one can obtain an upper limit on the mass of the halo
objects in any galaxy. Applying it to the Milky Way yields an upper limit
on the mass of dark matter clumps of $ M < 2 \times 10^6 M_{\odot} $.
Silk
and
Stebbins obtained, using slightly different arguments, a similar limit of
$ 10^6 M_{\odot} $ to 
avoid the problem of unacceptably heating the disk \cite{silk},
the general limit for halo objects so as not to heat the disk. However, if
the halo objects form pregalactically, which is the case if they are the
precusors to the galaxy,
then halo objects should have the same order of mass. Thus from an
analysis of
the gas-rich dwarf galaxy DD0154 one finds 
$ M \leq 7 \times 10^5 M_{\odot} $. For the dwarf galaxy 
GR8 the limit is $ M \leq 6 \times 10^3 M_{\odot} $.

Disk heating can also lead to `gravitational shocks'. During orbital
passage of a dark matter clump through the inner galaxy, the clump is
heated. This is a major factor in the disruption of globular star cluster
and hence is expected to be a primary factor in the disruption of cold
dark matter clumps \cite{salati}. This process is considered in the next
section.

\subsection{\it Dynamical Friction }

Gravitational shocking leads to considerable destruction of clumps.
The destruction timescale for clumps is given by \cite{salati}

\begin{equation}
\frac{ t_{dest} }{ t_{orbit} } \sim 4 \times 10^4
\left( \frac{10 pc }{ \left< r_{clump} \right> } \right)^2
\left( \frac{ R_{peri} }{ 8 kpc } \right)^2
\left( \frac{ v_{clump} }{ 100 km/s } \right)^2
\left( \frac{ 10^{10} M_{\odot} }{ M_{bulge } (R_{peri}) } \right)^2 
\left( \frac{ M_{clump} }{ 10^6 M_{\odot} } \right) 
\end{equation}

\noindent where 
$ M_{bulge} $ is the spherical or disk mass interior to the clump orbit,
\\
$ < r_{clump} > $ is the mean clump radius, \\
$ t_{orbit} $ is the internal orbit time,, \\
$ t_{dest} $ is the destruction time scale, \\
$ M_{\odot} $ is the solar mass, \\
$ M_{clump} $ is the mass of a clump, \\
$ v_{clump} $ is the mean clump velocity, \\
$ R_{peri} $ is the perigalactic distanve of the clump orbit. \\

Numerical simulations give a clump destruction rate of \cite{salati}

\begin{equation}
 \nu^{clump}_{destr} \sim 
   \left( \frac{ M_{gc } }{ M_{clump } } \right)^{1/3}
   \left( \frac{ \rho_{gc} }{ \rho_{clump} } \right)^{1/3}
   \nu^{gc}_{destr}
\end{equation}

where $ \nu^{gc}_{dest} $ is the present globular cluster destruction
rate, $ \rho_{gc} $ is the density of glonular clusters and $ M_{gc} $ 
their
mass.

The present globular destruction rate is obtained from numerical
simulations including the effect of
gravitational shocking by the disk and bulge and evaporation losses.
From such calculations it is inferred that this rate is $ 10^{-11} y^{-1}
$.

Thus, if the mass of the clump 
$ m_{clump } > 10^4 M_{\odot} $, clump
destruction is unimportant via dynamical processes over the last 
$ 10^{10} $ years \cite{salati}. 
Clumps in the inner few parsec of the halo are likely to
have been destroyed, while clump halos in the outer part of the galaxy
are
likely to have survived intact. If a fraction of $ f \sim 1 \% $ of the
halo has survived in clumps, and if 
$ \rho_{clump} \sim 10^3 \rho_{halo} $, 
then the resultant 
$ \gamma $ ray signal leads to 10 times larger
fluxes than for a uniform dark matter halo.

 Tormen, Diaferio and Syer \cite{tormen} also analysed the survival of
substructure in dark matter halos. Using high-resolution N-body
cosmological simulations, they analysed the survival of dark matter
satellites falling into larger bodies. They found that all satellites
preserve their identity for some time after merger. Satellites with less
than a fre percent mass of the halo may survive for several billion years,
whereas larger satellites rapidly sink into the center of the main halo
potential well and soon lose their identity. 

\subsection{\it Diffusion Towards the Galactic Center }

As halo objects traverse through the halo, they will lose energy and
consequently drift towards the Galactic nucleus. It has been shown
\cite{carr} that halo objects will be dragged into the nucleus by the
dynamical friction of the spherical stars from within a galactic radius

\begin{equation}
R_{df} = \left( M/10^6 M_{\odot} \right)^{2/3} 
         \left( t_g / 10^{10} y \right)^{2/3} kpc
\end{equation}

and that the total mass dragged into the Galactic nucleus is

\begin{equation}
M_{N} = 9 \times 10^8 
        \left( M/10^6 M_{\odot} \right)^2
        \left( \frac{t_g}{ 10^{10} y } \right)^2
        \left( \frac{a}{ 2kpc } \right)^{-2}
        \times M_{\odot}
\end{equation}

where $ a $ is the halo core radius and all other variables are as
defined above.

 This places a strong limit on the presence and survival of clumps.
However, there are severe caveats the this model. Once 2 clumps have
reached the center, they will form a binary, and should eventually
coalesce into a single body due to energy loss. If however, a third dark
matter clump arrives before coalescence occurs, then the `slingshot'
mechanism can eject one of the clumps and the remaining clumps could also
escape due to the recoil. Most clumps could escape eventual destruction by
this method, and hence this mehtod of destruction should not be
siginificant. In addition, dynamical friction will deplete the number of
stars in the nuclues, thereby suppressing dynamical friction.

Thus the cores of the clumps are expected to be
stable, although the outer layers may stripped off by the effects
given above. We conclude that a significant fraction of dark matter
probably survives in the form of clumps.

\subsection{\it Collisions }

Lake \cite{lake} analysed the stability of these dark matter clumps
against
collisions. The
timescale for disruption owing to mutual collisions of uniform-density
spheres yelds the critical value for survival versus destruction as :

\begin{equation}
\rho_c > \frac{ f^2 \rho_{h}^2 }{ 27 \rho_{crit} }
\end{equation}

\noindent where 
$ f $ is the fraction of dark matter in clumps with cluster masses $
M_{cl} \leq M \times 10^6 M_{\odot} $, 
$\rho_c $ is the density of WIMPs inside the clump,
$\rho_{h} $ is the density of halo dark matter, and
$ \rho_{crit} $ is the critical density of the Universe.

 If the density of the clump is larger than this critical value, then
collisions are not a major mechanism for disruption. This is since the
lower density clumps cannot survive collisions and are thus destroyed,
while the heavier clumps survive such collisions. Instead, 
when the density of the clumps satisfies this inequality, 
disruption by
tidal shocking becomes the dominant destruction mechanism of the clumps.
However, for reasonable parameters, the clumps survive disruption by this
mechanism also. Hence, clumps are expected to be stable against tidal
disruption.

\subsection{\it Experimental Searches for Dark Matter }

Clumping would have a significant impact on searches for dark matter. The
nondetection of dark matter candidates may be due to the lower value of
the local halo density. An observer in another part of the galaxy where
the local halo density was higher would observe a multitude of events that
would not be observable at other parts \cite{ira}. This holds for direct
detection methods. In such methods the dark matter is detected by
direct
interactions of the dark matter particle itself with the particles in the 
detector, analogous to the attempt to `capturing' it, or observing its
trail of escape.

Indirect detection techniques involve the 
search for signatures of the existence of dark matter rather than
attempting to directly capture it, such as
searching for the annihilation products of dark
matter particles.
Such methods 
would be facilitated by clumping, as a larger number of
such products would be observed on Earth in the direction of a clump
\cite{ira}. For example, the study of the high energy gamma-ray particles
produced due to the annihilation of dark matter particles will be much
more frequent in clumps of dark matter, and hence the excess prodution in
the direction of clumps can be calculated. These are constrained by the
results from experiments such as EGRET (to be explained below), which can
thus place limits on the
amount of clumping.


Recently, the neutralino has emerged as a more favourable dark matter
candidate. The discovery of a possible annual modulation signal in
recent
experimental searches for dark matter are compatible with the neutralino
as a dark matter candidate \cite{neutralino}.


\section{ Astrophysical Signatures }

\subsection{\it Dark Clumps }

 It has been observed that several collections of stars on the 
sub-galactic scale are dominated by dark matter. This may be taken as
already existing experimental evidence in favour of clumped dark matter.
At present, the results are generally approximated by `clumps' of dark
matter within a background uniform halo. Some of the recently
discovered dark-matter dominated clumps are :

\begin{itemize}
  \item Extreme dwarf spheroidal satellites of the Milky Way, Draco and
Ursa Minor are dominated by dark matter. 
They are essentially dark matter clumps.
Their existence is difficult to
reconcile with Gaussian initial conditions \cite{lake}.

  \item The Sagittarius
dwarf spheroidal, discovered in 1994, resides in the Milky Way halo, at a
distance of about 15 kpc from the centre of the Milky Way. It is situated
well inside the galactic halo potential and about 23 kpc from the Sun, and
is dark matter dominated \cite{lake}. 

\item R.C.Duncan analyzed the
absorption line
spectra of the double quasar Q2345+007 and found it to be lensed by a
low-luminosity object at z = 1.49 with a mass
contained within a volume of lateral extent $ \sim 40 h^{-1} kpc $
\cite{duncan}. This indicates
the existence of large clumps of dark matter in interstellar space.

\end{itemize}

V. Berezinsky, A. Bottino \& G. Mignola \cite{bere} showed that the recent
MACHOs discovered in the galaxy can be interpreted as dense neutralino
objects. However, the gamma-ray flux is many orders of magnitude higher
than the observed one \cite{bere}. Thus, the onserved gamma-ray flux
strongly constrians the fraction of dark matter that can have survived in
clumps in the case of the neutralino dark matter candidate.

\subsection{\it Gamma-Ray Sources }

 Clumps of dark matter could produce gamma-rays and act as gamma-ray
sources through the annihilation of WIMPs contained in the clumps of dark
matter. Lake also showed that the clumps would survive collisions with one
another and the disk \cite{lake}. He showed how several light sources in
the Cos B catalogue of gamma-ray sources could be candidates of gamma-ray
producing dark matter clumps. He proposed Geminga, the
second-brightest gamma-ray source above 50 MeV as a clump of dark matter.

\subsection{\it Gamma-Ray Halo }

  There is strong statistical support from EGRET (Energetic Gamma-Ray
Experiment Telescope) data for the existence of a gamma-ray halo
surrounding the galaxy. It has been suggested that this halo was the
result of annihilations of halo dark matter particles.
Bergstrom, Edsjo and P.Ullio set forth a model involving
moderately clumped neutralinos \cite{bergstrom}, \cite{clumpy}. 

  In this model the gamma rays are produced by the pair anihilation of
neutralinos $\Xi$ , the lightest supersymmetric particle in the Minimal
Supersymmetric Standard Model (MSSM). They assumed that the average clump
mass distribution follows the somooth component distribution. It was not
possible to reproduce the energy spectrum (antiproton flux vs. energy)
assuming a smooth dark matter halo. But it was compatible with the clumped
dark matter secnario.

A strong excess of gamma rays with
energy above 55 MeV was also detected towards the galactic
centre
of the Milky Way, This can be explained by the occurence of a higher
degree of clumping towards the galactic centre.
The degree of clumping required
implies a measurable excess of antiprotons at low enerigies. There is
support for this from recent BESS measurements.

 Thus the measured excess of cosmic gamma-rays and antiprotons could be
explained by invoking the existence of clumped dark matter. Bergstrom et
al pointed out upcoming experiments would be able to rule out or more
strongly confirm this theory.

 The integrated $ \gamma $ - ray flux above an energy threshold $ E_{th} $
is given by

\begin{equation}
 \phi_{\gamma} \left( E_{th}, \Delta\Omega, \psi \right) \sim
 1.87 \times 10^{-8} S \left( E_{th} \right) \times
 \left< J (\psi) \right> \left( \Delta\Omega \right)
 cm^{-2} s^{-1} sr^{-1}
\end{equation}

\noindent where $ S( E_{th} ) $ is a particle physics dependant term.
The dependance of the flux on the dark matter distribution is contained in
the term $ \left< J (\psi) \right> ( \Delta\Omega ) $.
This contribution of the clumps to the flux of gamma-rays, a measure ot
the contribution of clumps to the gamm-ray halo of the galaxy, is given
by \cite{indicate} :

\begin{equation}
\left< J ( \psi ) \right>_{cl} \left( \Delta \Omega \right) = \frac{1}{
8.5 kpc }
\frac{1}{ \Delta \Omega } f \delta 
\int_{\Delta \Omega} d \Omega^{`}
\int_{line of sight} dl \left( \frac{ \rho(l,\psi^{`}) }{ 0.3 GeV/cm^{3} }
\right)
\end{equation}

where $ \Delta \Omega $ is the angular acceptance of the detector pointing
in a direction which forms an angle $ \psi $ with respect to the galactic
center and $ f $ is the fraction of dark matter concentrated in clumps
and $ \rho(l, \psi) $ is the dark matter distribution at the point 
$ ( l, \psi ) $. In
the smooth case the dependance is quadratic in the density
$ \rho $. The relative strengths of the clumped components depends on both
the halo profile and the product of the halo fraction in clumps and their
overdensity, $ f \delta $.

 It is observed that there is a discrepancy between the 2.75 power law
expected from a model of diffuse galactic background and the results
obtained by EGRET. 
The dark matter signal carrying an excess of $ \gamma $ rays from a few
GeV to the mass of the dark matter particls $ M_{\chi} $ may explain this
discrepancy. In addition the smooth dark matter halo models cannot
reproduce the EGRET data, but this is possible with clumped dark matter
models.

The new high energy cosmic gamma-ray detectors both ground and space-based
shall have the possibility of searching for dark matter signals \cite{mw}.

\subsection{\it Antiproton Flux}

It has been suggested that the antiproton flux in cosmic rays may be due
to annihilating dark matter. In addition, the antiproton flux is able to
place constraints on the level of clumping as shown by Mitsui, Maki and
Orito 
\cite{mitsu}. The antiproton flux in cosmic rays may be due to
annihilating neutralino dark matter, evaporating primordial black holes
(PBH's) and superconducting strings. The annihilation of neutralinos may
produce a detectable flux of antiprotons. This process occurs via the path

\begin{equation}
\Xi \bar{\Xi} \rightarrow q, \,\ leptons \,\ and \,\
other \,\
particles
\rightarrow \bar{p} + other \,\ particles
\end{equation}

Using a standard model in the minimal $ N=1 $ supergravity with radiative
breaking of the electroweak gauge symmetry. Berezinsky 
{\it et al} \cite{omega}
pointed out that the expected value of $ \Omega_{\Xi} h^2 $ lies in the
range of $ 0.2 +- 0.1 $ in most cosmological models from the viewpoint of
the age of the Universe. Thus, Mitsui {\it et al } \cite{mitsu} took $
\Omega_{\Xi} h^2 =
0.18 $. Now, neutralinos with this value are most likely pure binos, which
predominantly annihilate into bottom quark pairs $ b \bar{b} $ or
tau lepton pairs $ \tau^{+} \tau^{-} $. The $ b \bar{b} $ pairs can give
antiprotons, but the tau leptons do not. Utilising a Monte Carlo
simulation based on the diffusion model utilising source spectra of $
\bar{p} $ ontained from the fragmentation functions constructed from
JETSET, Mitsui {\it et al }
found that the annihilation rate of neutralinos per unit volume
( $ S $ ) was proportional to the square of the local halo dark matter
density $ \rho_{\Xi} $, but that for PBH's, the relation was linear.

They also obtained fluxes of antiprotons that are too small if a
homogenous, isothermal and spherical dark matter distribution of density
$ \rho = 0.3 Gev/cm^3 $. They noted three casees, however, where the dark
matter density would be enhanced, leading to a larger antiproton flux.

Firstly, the Galactic halo may be flattened towards the galactic disk.
This could lead to an enhancement of the local halo density by a factor of
2, leading thereby to an enhancement in the antiproton flux by a factor of
4.

Secondly, an NGS (non-dissipative gravitational singularity) may reside at
the galactic centre, leading thereby to a halo distribution of 

\begin{equation}
\rho_{h}(r) = \rho_{h \odot } ( r / r_{\odot} )^{-1.8} \,\ for \,\ 
r > 0.1 pc
\end{equation}

where $ \rho_{h}(r) $ is the local halo density, 
$ r_{\odot} $ is the galactic radius of the Sun

and 
$ \rho_{h \odot } $ is the halo density near the Sun.
This would lead to an enhanced antiproton flux, possibly by a factor of
200.

Thirdly, if the halo consists of clumps of dark matter enhancements of $
10^2 - 10^9 $ may be generated in the early universe. If a few per cent of
the
neutralino dark matter is in such clumps, then the antipoton flux would be
enhanced by a factor of 20.

\section{ Geophysical Signatures }

Possible geophysical signatures of clumping include 
such large-scale
effects as volcanism and mass extinctions and the possible creation
of defects in mica.

\subsection{\bf{\it Volcanism }}

 Volcanism is one of the means through which heat generated inside the
Earth escapes it. The primary sources of heat in the Earth are believed to
be radioactivity and primordial accretional heat. As per conventional
wisdom these are the only major sources of heat in the Earth; in the case
of special planets like Io other sources like tidal heating can become
significant. Recently it has been shown by two of the authors \cite{vdm}
that dark matter accumulations inside the Earth could produce huge amount
of heat with drastic consequences. They proposed that the annihilation
heat due to dark matter could be a new source of heat.

Here the term `volcanogenic dark matter' is used to refer to
the volcanism arising due to
the capture and annihilation of dark matter inside the earth. This process
occurs via several stages: Firstly, the capture of dark matter, followed
by the accumulation of the captured dark matter in the centre of the Earth
and the annihilation of the same. This then leads to plume formation via
the process outlined below. The plume, on reaching the surface, leads to
massive volcanism of the flood basaltic type.

\subsubsection{\tt Capture of Dark Matter }

 The capture of dark matter particles by the Earth and planets was first
analysed by Press \& Spergel \cite{press}. This capture of dark matter
particles was investigated on account of the neutrinos and other
annihilation products produced. These products are currently the subject
of ongoing experiments \cite{berezin}. 
Krauss, Srednicki and Wilczek investigated the
neutrinos produced by captured dark matter particles, with scalar or Dirac
neutrinos as examples \cite{krauss}. They obtained greatly enhanced
capture rates for masses greater than $ 12 GeV $, and used the luminosity
of Uranus to place constraints on dark matter candidates. Upgoing muons
produced by accumulated dark matter inside Earth and Sun have also been
the subject of investigation \cite{bottino}.

Subsequently Gould obtained
greatly improved formulae with enhanced capture rates.
The Gould formula for the capture is \cite{gould} :

\begin{equation}
\dot{ N_{E} } = ( 4.0 \times 10^{16} sec^{-1} )
         \bar\rho_{0.4}
         \frac { \mu } { \mu_{+}^{2} }
         Q^{2}
         f
         \left< \hat\phi
                ( 1- \frac { 1-e^{-A^2} } { A^2 } )
                \xi_{1} (A)
                \right>
\end{equation}

\noindent where
$ \bar\rho_{0.4} $ is the halo WIMP density normalized to
   $ 0.4 GeVcm^{-3} $ ,
$ Q = N - ( 1 - 4 sin^2 \theta_W ) Z
   \sim N - 0.124Z $,
$ f $ is the fraction of the Earth's mass due to this element,
$ A^2 = ( 3 v^2 \mu ) / ( 2 \hat{v}^2 \bar\mu_{-} ) $,
$ \mu =  m_{X} / m_{N}  $,
$ \mu_{+} = ( \mu + 1 ) / 2  $,
$ \mu_{-} = ( \mu - 1 ) / 2  $,
$ \xi_{1} (A) $ is a correction factor,
$ v = $ escape velocity at the shell of Earth material ,
$ \hat{v} = 3kT_w/m_X
          = 300 kms^{-1} $ is the velocity dispersion, and
$ \hat\phi =  v^2 / { v_{esc} }^2  $
  is the dimensionless gravitational potential.

 This formula neglected the effect of the finite optical depth of the
Earth. If this effect is taken into account, multiple collisions occur,
and this enhances the capture by a factor of $ 5 - 30 \times $ 
\cite{arch}.
Since the Earth is in the potential well of the Sun, the WIMPs actually
move faster, making them more difficult to capture directly.
Gravitational diffusion however, leads to an increase in the 
phase-space density of free-space WIMPs due to encounters with the Earth,
Jupiter and Venus,
so that direct capture of bound and unbound WIMPs
is approximately equal to the case of direct capture in free space
\cite{grav}. This means that the formula given above holds good.
Gould showed that ndirect
capture, namely the capture of WIMPs that are in Solar orbits, was
insignificant \cite{grav}. Gould also obtained a high-precision
WIMP-capture formula for the Sun \cite{sun}.

\subsubsection{\tt Annihilation of Dark Matter }

 This dark matter that is captured gradually drifts towards the centre of the
Earth. This process occurs because the WIMPs lose energy.
Here they accumulate till the critical
threshold is crossed, beyond which annihilation becomes very important.
Above this threshold, virtually all the WIMPs captured are annihilated.
 In the WIMP mass range 15 GeV-100 GeV the Gould formula yields total
capture rates of the order of $ 10^{17} sec^{-1} $ to
$ 10^{18} sec^{-1} $ .
According to the Gould equation above, this yields
$Q_E \sim 10^8 W - 10^{10} W $ for a uniform density
distribution.

Interestingly intriguing signatures of a dark matter WIMP have been
recently obtained at 56 GeV \cite{60}. This mass of WIMP, nearly the same
as that of iron, would lead to a resonant enhancement in the capture of
WIMPs by Earth. The consequence would be an increase in the heat
production
of the Earth, and this would be in support of our model calculations.

\subsubsection{\tt Creation of Plume }

 In case of uniformly distributed dark matter, the dark matter heat
production is much less than the annual geothermic heat output of
$ 4 \times 10^{12} W $. However, during the passage of a clump core, the heat
production due to annihilating dark matter exceeds that due to other
sources by several orders of magnitude. What would be the consequence of
these vast amounts of heat ?

 As per conventional theory, this would lead to the creation of what are
referred to as `deep mantle plumes'. As per standard geothermologic models,
the lowest layer of the mantle ( called the D" layer) absorbs heat from the
core, thereby decreasing in density. Eventually, once a critical minimum
density is crossed, this D" layer breaks up into rising plumes. These plumes
are an efficient way of heat transfer. Ultimately, the largest plumes are
capable of reaching the surface.

\subsubsection{\tt Flood Basalt Volcanism and Geomagnetic Reversals }

 On arrival at the surface, the plume will melt its way through the crust,
leading to initial explosive volcanism (due to the molten crust) followed by
a much longer period of flood basaltic volcanism. As is observed from the
vast relics left behind by these episodes, such volcanism is known to be the
most most extensive form of volcanism in the world. Examples are the massive
Deccan flood basalt volcanic province, the Siberian flood basalt volcanic
region, and the Brazilian flood basalt zone. Many such flood basalt episodes
occur simulatenous to major mass exinctions of life. Much evidence of a link
between these volcanic episodes and the concordant mass extinctions has been
established \cite{vdm}.

 In addition, the large quantities of heat produced inside the core are
capable of leading to geomagnetic reversals. The rise of temperature inside
the core leads to an instability in the generation of the magnetic field,
which may lead to geomagnetic reversals.

 A very attractive feature of this model is that it can explain the 30
million year periodicity in the record of mass extinctions, the 30 million
year periodicity observed in the periodicity of flood basalt volcanism, and
the 30 million year periodicity calculated for the crossing of clumps
(obtained from the models of clumping).

\subsection{ Defects in Mica }

Baltz, Westphal and Snowdon-Ifft studied the effect of dark matter on
mica \cite{mica}. They studied two models of the density profiles for the
clumped fraction of the dark matter halo, namely the Hernquist profile and
the Navarro, French and White (NFW) profiles. The Hernquist profile is
given by:

\begin{eqnarray}
 \rho(r) = \frac{M}{2 \pi } \frac{a}{ r(r + a )^3 }, \\
 \phi(r) = - \frac{ GM }{ r + a }
\end{eqnarray}

 The NFW profile is given by 

\begin{eqnarray}
 \rho(r) = \frac{M}{4 \pi} \frac{1}{ r(r+a)^2 }, \\
 \phi(r) = - \frac{GM}{ r } ln \left( 1 + \frac{r}{a} \right)
\end{eqnarray}

\noindent where $ a $ is the scale radius, 
$ M $ the mass parameter, 
$ \rho $ the density potential,
$ \phi $ the gravitational potential,
and $ r $ the radius at the point under consideration.

Now, the rate at which defects accumulate in mica is a function of the
angle $ \alpha $ which the incident particle makes with the cleavage
plane. Monte Carlo simulations show that the rate at which tracks
accumulate, $ \frac{ dN }{ dt } $ is well approximated by \cite{mica} :

\begin{equation}
 \frac{ dN ( \alpha ) }{ dt } = c_{0} + c_{1} | cos \alpha |
\end{equation}

where
$ c_0, c_1 $ are functions of the WIMP mass, the dispersion velocity of
the halo and the velocity of the earth
through
the halo. They were calculated assuming the mica has remained fixed in
position relative to the incident WIMPs. In geophysical terms, this
corresponds to assuming that the mica has remained geologically
stationary, in other words the phenomenon of continental drift has been
ignored. They took, as an example, a single $ 10^6 M_{\odot} $ clump
interacting with the earth at different times from the formation of the
mica, taken to be 440 million years ago, and the present.
 Experiments using neutrons instead of WIMPs as the incident particles
conformed to this model, and hence it is reasonable.

Baltz, Westphal and Snowdon-Ifft took for their simulations, a mica age of
440 million years, which corresponds to two galactic years, ie. two
rotations of the earth about the galactic centre. For A=10, where A is 
the mass number, they noticed a
substantial change in the signal of the clump compared to the background.
At A=100, the change in the signal is much more substantial. Most of the
forward angle show a pronounced change in the signal. In the clumped case
the signal is much larger and the maximum and minimum directions are very
different from the non-clumped case. For A=1000, the signal is ten times
larger than the non-clumped case \cite{mica}. The signal from a clump
encounter 440 My ago appears to be oriented in the same direction for all
incident directions. This circumstance arises since the North American
craton was situated at the equator at this time. For signals occurring 220
My ago, this degenracy is broken as this continent had by that time moved
away from the equator. The authors suggested using mica samples from
different locations from different continents, of different ages and
different orientations to obtain a large amount of information about the
nature of the clumping of the dark matter in the halo \cite{mica}.

\newpage
\section{ Conclusion }

In addition to arising naturally within cosmological dark matter
modles, clumped dark matter would be stable on astrophysical time scales
and clumps are likely to have survived to the present epoch. The existence
of
clumped dark matter also would solve many problems in astrophysics and
geology that find either difficult or no explanation at all within
conventional
frameworks. Such cases include the galactic gamma-ray halo and the
periodicity in terrestrial flood basalt volcanism. 
Annihilation of clumped dark matter should also be treated as a new source
of heat in planetary bodies which may have observable consequences.
In addition they may
have
already been discovered in the form of dwarf spheroidals.
The significance of clumped dark matter is such that it calls for further
studies.

\end{document}